\documentclass[aip, pop, amsmath, amssymb, reprint]{revtex4-1}

\usepackage{graphicx}% Include figure files
\usepackage{dcolumn}% Align table columns on decimal point
\usepackage{bm}% bold math

\begin{document}

\preprint{AIP/123-QED}

\title[Properties of magnetized Coulomb crystals of ions with polarizable electron background.]
{Properties of magnetized Coulomb crystals of ions with polarizable electron background}

\author{A.A.\, Kozhberov}
\email{kozhberov@gmail.com}

\affiliation{Ioffe Institute, Politekhnicheskaya 26,
St. Petersburg, 194021, Russia}%

\date{\today}

\begin{abstract}
We have studied phonon and thermodynamic properties of a body-centered cubic (bcc) Coulomb crystal of ions with weakly polarized electron background in a uniform magnetic field ${\bf B}$. At $B=0$ the difference between phonon moments calculated using the Thomas-Fermi (TF) and random phase (RPA) approximations is always less than one percent and for description of phonon properties of a crystal TF formalism was used. This formalism was successfully applied to investigate thermodynamic properties of magnetized Coulomb crystals. It was shown that the influence of the polarization of the electron background is significant only at $\kappa_{\rm TF}a > 0.1$ and $T \ll T_{\rm p}(1+h^2)^{-1/2}$, where $\kappa_{\rm TF}$ is the Thomas-Fermi wavenumber, $a$ is the ion sphere radius, $T_{\rm p}\equiv\hbar \omega_{\rm p}$ is the ion plasma temperature, $h\equiv \omega_B/\omega_{\rm p}$, $\omega_B$ is the ion cyclotron frequency and $\omega_{\rm p}$ is the ion plasma frequency.
\end{abstract}

\maketitle

\section{Introduction}
\label{intro}
A system of point charged particles (ions, electrons, grains of dust) arranged in a lattice and immersed into a charge-neutralizing background is called a Coulomb crystal. This model is in use in the theory of dusty plasma (e.g., \citet{F}), in solid state physics (e.g., \citet{BH54}) and in the theory of degenerate stars. The last application is of the greatest interest for us. The matter in neutron star crusts and cores of old white dwarf consists of fully ionized atoms and degenerate electrons. It could be successfully described as a Coulomb crystal of ions (e.g., \citet{ST83,HPY07}).

For many practical purposes it is a good approximation to consider the electron background in the Coulomb crystal as constant and uniform. However, for more precise calculations, the effect of electron charge screening should be taken into account, especially at low densities (e.g., \citet{B02}).

In the present paper we consider the influence of the electron background polarizability on the Coulomb crystals of ions in magnetic fields. This investigation could be important for understanding properties of magnetars, neutron stars with an extremely high magnetic field (up to $10^{15}$ G on the surface). Previously, magnetized Coulomb crystals with uniform electron background were extensively studied (e.g., \citet{UGU80,NF82,NF83,B09,BY13}). Coulomb crystals with polarizable electron background in a uniform magnetic field were discussed, for the first time, in our recent paper (\citet{BK17}), where their phonon properties were described. In the present paper we calculate phonon frequency moments and analyze thermodynamic properties of such systems.

\section{Coulomb crystals without magnetic field}
The electron charge screening could be described by a dielectric function. We assume that ion motion is slow while the electron density response to this motion is instantaneous. This assumption allows us to use the static longitudinal dielectric function $\epsilon(q)$, where $q$ is the length of a wavevector.

Firstly, consider a Coulomb crystal without magnetic field. The dielectric function of nonmagnetized degenerate relativistic electron gas based on the random-phase approximation was obtained by \citet{J62} (RPA approach):
\begin{eqnarray}
\epsilon(q)&=&1+\frac{\kappa^2_{\rm TF}}{q^2}\left\{\frac{2}{3}-\frac{2}{3}\frac{y^2_{\rm r}x_{\rm r}}{\gamma_{\rm r}}\ln\left(x_{\rm r}+\gamma_{\rm r}\right)\right. \nonumber \\
&+&\frac{x^2_{\rm r}+1-3x^2_{\rm r} y^2_{\rm r}}{6y_{\rm r}x^2_{\rm r}}\ln\left|\frac{1+y_{\rm r}}{1-y_{\rm r}}\right| \label{UJan} \\
&+&\left. \frac{2y^2_{\rm r} x^2_{\rm r} -1}{6y_{\rm r} x^2_{\rm r}}\frac{\sqrt{1+x^2_{\rm r} y^2_{\rm r}}}{\gamma_{\rm r}}\ln \left|\frac{y_{\rm r}\gamma_{\rm r}+\sqrt{1+x^2_{\rm r} y^2_{\rm r}}}{y_{\rm r}\gamma_{\rm r}-\sqrt{1+x^2_{\rm r} y^2_{\rm r}}}\right|\right\}~, \nonumber
\end{eqnarray}
where $\gamma_{\rm r}\equiv\sqrt{1+x_{\rm r}^2}$, $x_{\rm r}\equiv p_{\rm F}/(m_e c)$ is the electron relativity parameter, $y_{\rm r}=q/(2 p_{\rm F})$, $\kappa_{\rm TF} \equiv \sqrt{4 \pi e^2 \partial n_e/\partial\mu_e}$ is the Thomas-Fermi wave number, $e$, $m_e$, $n_e=Zn$, and $\mu_e$ are the electron charge modulus, mass, number density, and chemical potential, respectively, $p_{\rm F}$ is the electron Fermi momentum, $n$ and $Z$ are number density and charge number of ions.

For a strongly degenerate electron gas, $\epsilon(q)$ could be considered as a function of $q$ and two independent parameters of the system:
\begin{equation}
\kappa_{\rm TF}a\approx0.185Z^{1/3}\frac{(1+x_{\rm r}^2)^{1/4}}{x_{\rm r}^{1/2}}
\label{ktf}
\end{equation}
and $Z$ or $x_{\rm r}\approx 1.00884(\rho_6 Z/A)^{1/3}$, where $A$ is mass number of ions, $a \equiv (4 \pi n /3)^{-1/3}$ is the ion sphere radius, $\rho_6\equiv \rho/10^6$ g cm$^{-3}$.

In the long-wavelength limit ($q \ll  p_{\rm F}$), the dielectric function reduces to the well-known Thomas-Fermi form (TF approach):
\begin{equation}
\epsilon(q)=1+\frac{\kappa^2_{\rm TF}}{q^2}.
\label{UJan2}
\end{equation}
In this case $\epsilon(q)$ could be considered as a function of $q$ and $\kappa_{\rm TF}a$. If the electrons are fully degenerate, equations (\ref{UJan}) and (\ref{UJan2}) are valid if the polarization of the electron background is weak $\kappa_{\rm TF}a \leq 1$. In the  ultrarelativistic limit $\kappa_{\rm TF}a\approx0.185Z^{1/3}$. $\kappa_{\rm TF}a$ decreases with increasing $\rho$ hence the value $\kappa_{\rm TF}a=0.185Z^{1/3}$ should be considered as a lower limit. For fully ionized iron $^{56}$Fe it is equal to $0.548$. Furthermore, for $^{56}$Fe ions, $\kappa_{\rm TF}a=1$ at $\rho\approx6.6 \times 10^4$ g cm$^{-3}$.

It is also interesting to note that our model of the Coulomb crystal of ions with polarized electron background is similar to the strongly-coupled Yukawa system which was described in \citet{HF94} with only one difference. In our model electrons are strongly degenerated while in \citet{HF94} the background is nondegenerate gas of ions and electrons.

In the absence of the magnetic field, ion equation of motion in the crystal can be written as (e.g., \citet{BK17})
\begin{eqnarray}
    \omega^2_{{\bm k}\nu} A_{{\bm k}\alpha} &=&
    D_{\alpha \beta}({\bm k}) A_{{\bm k}\beta}  \\
    D_{\alpha \beta}({\bm k}) &=& \frac{\omega_{\rm p}^2}{(2 \pi)^3n}
    \frac{\partial^2}{\partial X_\alpha \partial X_\beta}
    \sum_{I \ne 0}
    \left(1- e^{-i{\bm k} \cdot {\bm R}_I}\right) \nonumber \\
    &\times& \int \frac{{\rm d}{\bm q}}{q^2 \epsilon(q)}
    e^{i{\bm q} \cdot ({\bm R}_I-{\bm X})} \bigg\vert_{{\bm X}=0} \label{sle}~,
\end{eqnarray}
where $\omega_{\rm p}=\sqrt{4\pi n Z^2 e^2/M}$ is the ion plasma frequency, $M$ is the ion mass, ${\bm A}_{\bm k} = \sum_I {\bm u}_I e^{-i{\bm k}\cdot {\bm R}_I}$. Dynamic matrix $D_{\alpha \beta}({\bm k})$ is determined by the quadratic term in the expansion of the potential energy in a series over powers of ion displacements ${\bm u}_I$ about the equilibrium positions ${\bm R}_I$. Equation (\ref{sle}) is appropriate for any $\epsilon(q)$ and any lattice with one ion in the elementary cell (like the body-centered cubic lattice).

Phonon spectrum of the body-centered cubic (bcc) lattice consists of three modes (index $\nu=1 \dots 3$). For nonmagnetized crystal it was studied in \citet{B02} where it was shown that modes appear to be virtually the same in the RPA and TF approaches for all wave vectors ${\bm k}$. In other words $\omega_{{\bm k}\nu}$ is virtually independent of $x_{\rm r}$. This statement could be illustrated by phonon frequency moments: $u_1 \equiv \langle\omega_{{\bm k}\nu}/\omega_{\rm p}\rangle$ and $u_{-1} \equiv \langle\omega_{\rm p}/\omega_{{\bm k}\nu}\rangle$, where $\langle\dots\rangle$ denotes the averaging over all modes in the first Brillouin zone of the lattice:
\begin{equation}
\langle f_{{\bm k}\nu}\rangle=\frac{(2\pi)^3 n}{3} \sum_{\nu=1}^{3}\int\textrm{d}{\bm k}f_{{\bm k}\nu}~.\
\end{equation}

\begin{table}
\caption{\label{tab:1} Moments of the phonon spectrum bcc lattice.}
\begin{ruledtabular}
\begin{tabular}{ccc}
$x_{\rm r}$ & $u_1$ & $u_{-1}$ \\
\hline
\multicolumn{3}{c}{TF approach} \\
 & 0.4852393 & 2.90710 \\
\hline
\multicolumn{3}{c}{RPA approach} \\
10 & 0.4864944 & 2.89936 \\
3  & 0.4861845 & 2.90118 \\
1  & 0.4858736 & 2.90313 \\
0.3 & 0.4861221 & 2.90100 \\
0.1 & 0.4876263 & 2.88325 \\
\end{tabular}
\end{ruledtabular}
\end{table}
Moments of the bcc lattice at $\kappa_{\rm TF}a=0.5$ are presented in Table \ref{tab:1}. For both cases we use the same integration grid. One can see that dependence of $u_1$ and $u_{-1}$ on $x_{\rm r}$ is weak and non-monotonic. Note that the electrostatic energy has the same non-monotonic behavior\cite{B02}. The difference between RPA and TF approaches is small and does not exceed 1\% but it could not be always neglected.
The first moment determine the energy of zero-point vibrations $E_0 \equiv 1.5N\hbar \omega_{\rm p} u_1$. For the face-centered cubic (fcc) lattice $u_1=0.4871003$ at $\kappa_{\rm TF}a=0.5$ in TF case. So the difference between $u_1$ of the bcc and fcc lattices in TF approach is also $\lesssim 1$\%.
Hence if we would like to calculate which lattice has the smallest total free energy we should use the RPA formalism for a more accurate result\cite{K16}.

On the other hand, the TF approach makes possible to calculate the phonon and thermodynamic properties of the Coulomb crystal with an error of the order of 1 percent in comparison with the more accurate method. This accuracy is sufficient to describe the neutron star and white dwarf envelopes.

\section{Coulomb crystals in magnetic field}
\begin{figure}
\center{\includegraphics[width=1.0\linewidth]{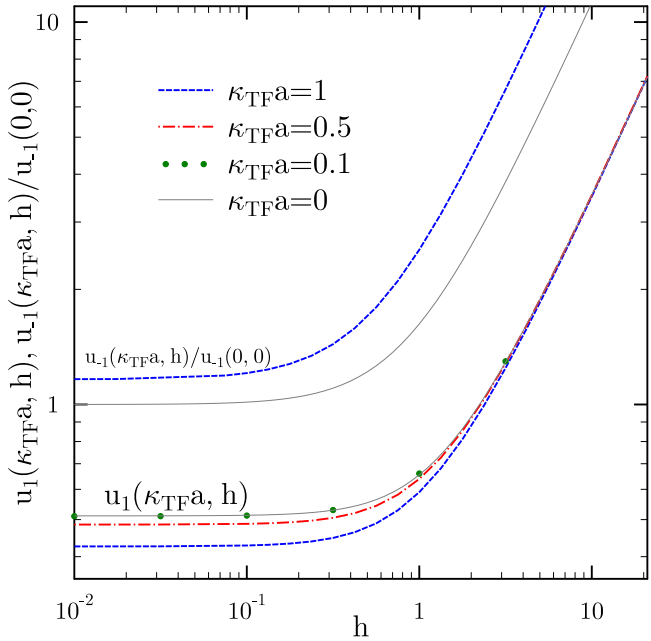}}
\caption{Dependence of $u_1(\kappa_{\rm TF}a,h)$ and $u_{-1}(\kappa_{\rm TF}a,h)$ on $h$ for different $\kappa_{\rm TF}a$.}
\label{fig:mom_pol_h}
\end{figure}
Consider a Coulomb crystal with polarizable electron background in the uniform magnetic field ${\bm B} \equiv {\bm n} B$, where ${\bm n}$ is the unit vector in the direction of the field. The general formulae for the dielectric function of a degenerate magnetized relativistic electron gas based on the random-phase approximation has a rather cumbersome form\cite{ST62}. In accordance with the previous section we restrict ourselves to the TF approach: $\epsilon(q) = 1 + \kappa_{\rm TF}^2/q^2$. In the presence of the magnetic field and at $T=0$  the Thomas-Fermi wave number is equal\cite{HPY07}:
\begin{equation}
   \kappa_{\rm TF}^2 = \frac{2Be^3\mu_e}{\pi c^2 \hbar^2}
   \sum_{l=0}^{l_{\max}} \frac{2-\delta_{l,0}}{\sqrt{\mu_e^2-(m_e c^2)^2-2\hbar ceBl}}~,
\end{equation}
where $l$ is the Landau level number. $\kappa_{\rm TF}$ oscillate around its value obtained neglecting the magnetic field (Eq. (\ref{ktf})). The anomalous magnetic moment of electrons is ignored.

Ion oscillation equation for crystal in the uniform magnetic field could be rewritten as\cite{BK17}
\begin{equation}
      D_{\alpha \beta}({\bm k}) A_{{\bm k}\beta}
      - \Omega^2_{{\bm k}\nu} A_{{\bm k}\alpha}
      - i \Omega_{{\bm k}\nu} \omega_B \varepsilon_{\alpha \beta \gamma} n_\beta A_{{\bm k}\gamma} = 0~,
      \label{mag}
\end{equation}
where $\omega_B=ZeB/(Mc)$ is the ion cyclotron frequency and $\varepsilon_{\alpha \beta \gamma}$ is the Levi-Civita symbol. The dynamic matrix $D_{\alpha \beta}({\bm k})$  is the same as in Eq. (\ref{sle}). The ion vibration frequencies $\Omega_{{\bm k}\nu}$ depend on ${\bm k}$, $\kappa_{\rm TF}$, $B$ and ${\bm n}$.

It is thought that the magnetic field in the crystal is directed with respect to the crystallographic axes in such a way that the zero-point energy is minimized\cite{UGU80}. In the bcc lattice with rigid electron background, $E_0$ is minimum if magnetic field is directed towards one of the nearest neighbors\cite{B09}, for example, ${\bm n}_{\rm{min}}=(1,1,1)/\sqrt{3}$. The polarization of the electron background does not change this direction, ${\bm n}_{\rm{min}}$ stays equal $(1,1,1)/\sqrt{3}$ at any $\kappa_{\rm TF}$. At fixed $\kappa_{\rm TF}$, the dependence of $u_1$ on ${\bm n}$ is weak. The difference between the minimum and maximum values of $u_1$ does not exceed 1\% similar to the uniform background case\cite{B09,K16}.
So we can fix ${\bm n}={\bm n}_{\rm{min}}$ and consider dependence of moments of the phonon spectrum and phonon thermodynamic functions only on $\kappa_{\rm TF}a$ and $h\equiv \omega_B/\omega_{\rm p}=B/(c\sqrt{4\pi \rho})\approx0.941 B_{15}/\sqrt{\rho_8}$, where $B_{15}=B/10^{15}$ Gauss and $\rho_8=\rho/10^8$ g cm$^{-3}$. Hence in magnetars with internal crustal magnetic field $10^{15}$ G $h>1$ at $\rho \lesssim 8.85 \times 10^7$ g cm$^{-3}$.

In Fig. \ref{fig:mom_pol_h} we plot the dependence of $u_1(\kappa_{\rm TF}a,h)$ on $h$ for different $\kappa_{\rm TF}a$. In the chosen scale curves for $\kappa_{\rm TF}a=0.1$ and 0.0 are indistinguishable. For instance, $u_1(0,0.1)=0.513239$ and $u_1(0.1,0.1)=0.512234$. At $h\ll1$ the magnetic field could be neglected and all moments tend to constant. This limit was studied in \citet{B02}. The influence of  the electron charge screening on $u_1$ decreases with increase of the magnetic field. At $h\gg1$ the main contribution to the first moment comes from the highest cyclotron mode $\Omega_{{\bm k}3}$ which does not depend on $\kappa_{\rm TF}a$. Hence $u_1(\kappa_{\rm TF}a,h)\approx h/3$ at $h\gg1$.

Ratio $u_{-1}(\kappa_{\rm TF}a,h)/u_{-1}(0,0)$ is shown in Fig. \ref{fig:mom_pol_h} for $\kappa_{\rm TF}a=1$ and $\kappa_{\rm TF}a=0$, where $u_{-1}(0,0)=2.79853$. The lowest mode of the bcc lattice spectrum $\Omega_{{\bm k}1}\propto 1/h$ near the center of the first Brillouin zone but it also has a complicated\cite{BK17} dependence on $\kappa_{\rm TF}a$. Hence at $h\gg1$ the moment $u_{-1}(\kappa_{\rm TF}a,h)$ depends on $h$ and $\kappa_{\rm TF}a$ and for $u_{-1}(\kappa_{\rm TF}a,h)$ polarization effects are important at any $h$.

\section{Thermodynamic functions of Coulomb crystals in magnetic field}
\begin{figure}
\center{\includegraphics[width=1.0\linewidth]{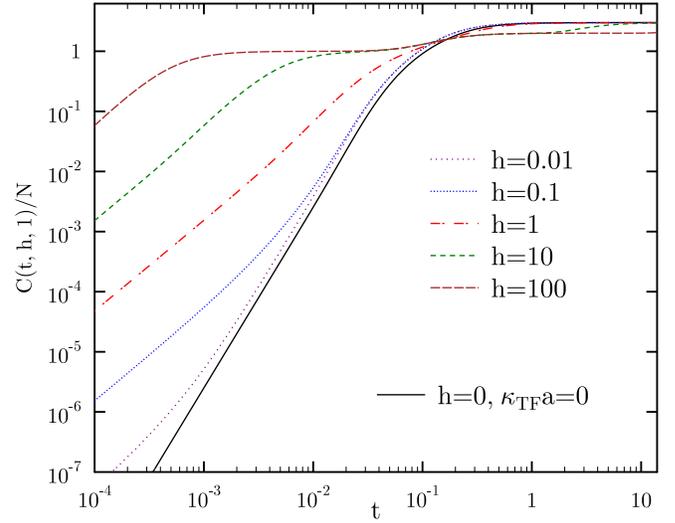}}
\caption{Dependence of $C(t,h,1)/N$ on $t$ for different $h$.}
\label{fig:td_pol_h_1}
\end{figure}

\begin{figure}
\center{\includegraphics[width=1.0\linewidth]{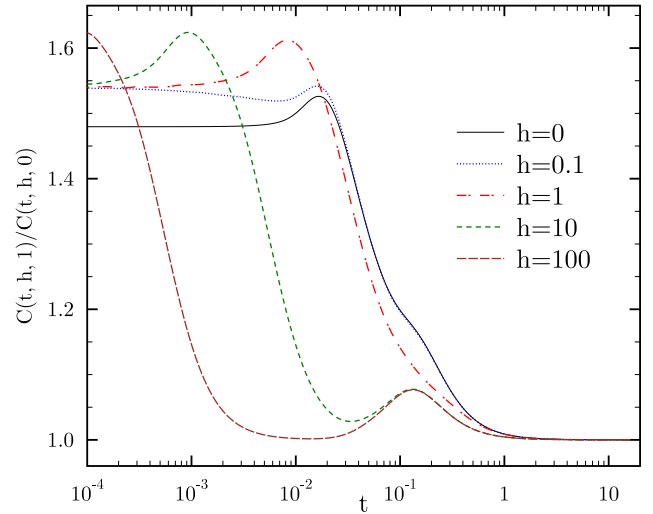}}
\caption{Dependence of $C(t,h,1)/C(t,h,0)$ on $t$ for different $h$.}
\label{fig:td_pol_h_2}
\end{figure}
Among all thermodynamic functions of Coulomb crystals with polarized electron background in magnetic field we consider the phonon heat capacity $C(t,h,\kappa_{\rm TF}a)=3N\langle w^2/(4 \sinh^2(w/2))\rangle$ and the Helmholtz free energy $F(t,h,\kappa_{\rm TF}a)=3NT\langle \ln (1-e^{-w})\rangle$, where $w\equiv \hbar \Omega_{{\bm k}\nu}/T$, $t\equiv T/T_{\rm p}$ and $T_{\rm p}\equiv\hbar\omega_{\rm p}$.

Polarization of the electron background tends to increase the heat capacity but at $\kappa_{\rm TF}a=0.1$ ratio $C(t,h,0.1)/C(t,h,0)$ lays between 1 and 1.0055 at any $t$ and $h$. Thus at such small $\kappa_{\rm TF}a$ polarization effects on the thermodynamic properties could be neglected.

In Fig. \ref{fig:td_pol_h_1} dependence of $C(t,h,1)/N$ on $t$ for different $h$ is plotted. At $h\gg1$ frequencies in the spectrum differ from each other by orders of magnitude. Therefore, at $t\gg1$ in the temperature dependence of the heat capacity three plateau appear: $C=N$, $C=2N$ and $C=3N$, when contributions come from the lowest mode in the entire Brillouin zone, from two lowest modes and from all modes, respectively. Near the center of the first Brillouin zone the lowest mode $\Omega_{{\bm k}1}\propto k^2/h$. This mode makes a major contribution to the thermodynamic functions at low temperatures. Hence $C(t,h,\kappa_{\rm TF}a)\propto (th)^{3/2}$ at $t\ll(1+h^2)^{-1/2}$. Dependence of the phonon spectrum on $\kappa_{\rm TF}a$ is rather complicated and the behavior of the thermodynamic functions with $\kappa_{\rm TF}a$ can not be described so easily.

\begin{figure}
\center{\includegraphics[width=0.990\linewidth]{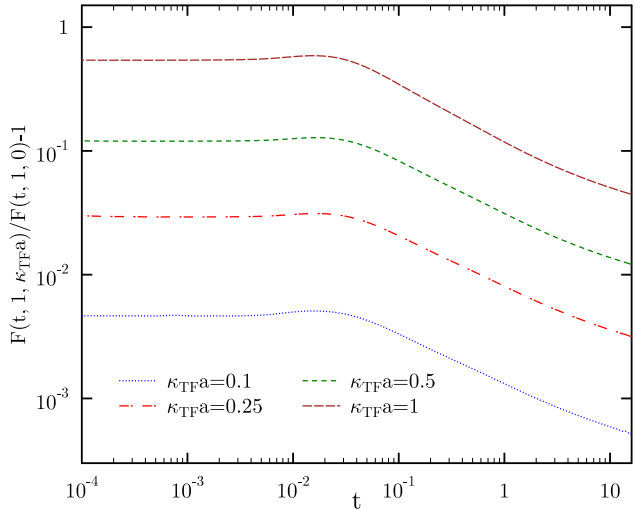}}
\caption{Dependence of $F(t,1,\kappa_{\rm TF}a)/F(t,1,0)-1$ on $t$ for different $\kappa_{\rm TF}a$.}
\label{fig:td_pol_h_3}
\end{figure}
The effect of the electron charge screening is illustrated in Fig. \ref{fig:td_pol_h_2}, where dependence of $C(t,h,1)/C(t,h,0)$ on $t$ for different $h$ is plotted. Polarization of the electron background significantly changes modes only near the center of the first Brillouin zone. That's why $C(t,h,1)/C(t,h,0)$ becomes noticeably greater than 1 only at low temperatures. The magnetic field reduces the effect of the polarization on the heat capacity. Influence of the polarization shifts to the lower temperatures with increasing $h$. At $t\sqrt{1+h^2}\approx1.4\times10^{-2}$ ratio $C(t,h,1)/C(t,h,0)$ reaches maximum. For example, at $h=1$ it is equal $1.61$. At $t\ll(1+h^2)^{-1/2}$\,\,\,$C(t,h,1)/C(t,h,0) \approx 1.54$ at any $h\gg0$.

Relative difference between the thermal contributions to the free energy of different $\kappa_{\rm TF}a$ and $\kappa_{\rm TF}a=0$ is plotted in Fig. \ref{fig:td_pol_h_3}. At $t\ll 1$ ratio $F(t,1,\kappa_{\rm TF}a)/F(t,1,0)$ does not depend on temperature. At low temperatures and $\kappa_{\rm TF}a \ll 1$ it is roughly sufficient to write that $F(t,1,\kappa_{\rm TF}a)/F(t,1,0)-1\propto (\kappa_{\rm TF}a)^2$.
At $t\approx0.016$ ratio $F(t,1,\kappa_{\rm TF}a)/F(t,1,0)$ reaches a rather weak maximum. For example, $F(0.016,1,0.5)/F(0.016,1,0)\approx1.1279$, while at $t\rightarrow 0$ $F(t,1,0.5)/F(t,1,0)\approx1.1198$. At high temperatures $F(t,h,\kappa_{\rm TF}a)$ does not depend on $h$ according to the Bohr van Leeuwen theorem.

As shown in \citet{K16} at $\kappa_{\rm TF}a=0$ the thermal contribution to the total Helmholtz free energy is important only at $t/\sqrt{1+h^2} \gg 0.01$. A similar situation takes place at $\kappa_{\rm TF}a>0$. For instance, at $\kappa_{\rm TF}a=1$ and $h=1$ thermal contribution to the total Helmholtz free energy is noticeable only at $t\gtrsim 0.02$.

Thus, the polarization of the electron background leads to an increase of the heat capacity of the magnetized neutron star crust but only at low temperatures. At the same time this polarization can make a noticeable  contribution to the total free energy of the crystal at $t/\sqrt{1+h^2} \gg 0.01$ and $\kappa_{\rm TF}a \sim 1$.

\section{Conclusions}
It was shown that the difference between phonon frequency moments of a Coulomb crystal without magnetic field calculated using the Thomas-Fermi and random phase approximations is less than one percent. This difference could not be neglected when the total energies of different lattices are compared but usually such small difference is not important for investigations.

The Thomas-Fermi formalism has been used to study phonon moments and thermodynamic functions of magnetized Coulomb crystals of ions with polarizable electron background. For $u_{-1}$ influence of the polarization of the electron background is important at any $h$, but $u_1$ does not depend on $\kappa_{\rm TF}a$ at $h \gg 1$. Polarization of the electron background tends to increase the heat capacity at low temperatures $t \ll (1+h^2)^{-1/2}$ but even at $\kappa_{\rm TF}a=1$ the ratio $C(t,h,1)/C(t,h,0)$ is always less than 1.63.

These results could be used for realistic calculations of the thermal evolution of neutron star crusts and white dwarf cores.

\begin{acknowledgments}
The author is deeply grateful to D.A. Baiko and D.G. Yakovlev for discussions.
This work was supported by Russian Science Foundation grant 14-12-00316.
\end{acknowledgments}

\end{document}